\documentclass[twocolumn,showpacs,amsmath,prl]{revtex4-1}

\usepackage{amsfonts}
\usepackage{amssymb}
\usepackage{graphicx}
\usepackage{dcolumn}
\usepackage{bm}
\begin{document}

\title{Antiferromagnetic crystalline topological insulators}

\author{Chao-Xing Liu$^1$}

\affiliation{ $^1$ Department of Physics, The Pennsylvania State University, University Park, Pennsylvania 16802-6300
}

\date{\today}

\begin{abstract}
The gapless surface Dirac cone of time reversal invariant topological insulators is protected by time reversal symmetry due to the Kramers' theorem. Spin degree of freedom is usually required since Kramers' theorem only guarantees double degeneracy for spinful fermions, but not for spinless fermions. In this paper, we present an antiferromagnetic spinless model, which breaks time reversal symmetry. Similar to time reversal invariant topological insulators, this model possesses a topologically non-trivial phase with a single surface Dirac cone, which is protected by the combination of time reversal and translation operation. Our results show that in magnetic crystals, a single Dirac cone can exist on the surface even without any spin degree of freedom and spin-orbit coupling. 
\end{abstract}

\maketitle
Time reversal (TR) invariant topological insulators (TIs) \cite{qi2010a,moore2009,hasan2010,qi2011} possess insulating bulk states, and metallic surface states consisted of an odd number of Dirac cones, e. g. a single Dirac cone in the prototype TIs, Bi$_2$Se$_3$ family of materials \cite{hjzhang2009,chen2009,xia2009}. The gapless nature of topological surface states is protected by TR symmetry according to the Kramers' theorem\cite{kane2005a,bernevig2006a,kane2005b,fu2007a}. For a TR invariant system, TR operation $\Theta$ satisfies $\Theta^2=-1$ for spinful fermions and $\Theta^2=1$ for spinless fermions or bosons. According to the Kramers' theorem, when $\Theta^2=-1$ there is a double degeneracy for each energy level, which protects the gapless nature of surface Dirac cones in TR invariant TIs. Therefore, spin degree of freedom, as well as spin-orbit coupling, is required for TR invariant TIs.  

The idea of topological phases due to the symmetry protected degeneracy is generalized to other types of symmetry, such as crystalline symmetry\cite{fu2011,hsieh2012,dziawa2012,tanaka2012,xu2012,slager2012,fang2012,fang2013,jadaun2012}, particle-hole symmetry\cite{qi2009b,sato2009,schnyder2008}, {\it etc}. In this work, we present a spinless antiferromagnetic model with topologically non-trivial phases, which breaks TR symmetry, but possesses surface states of a single Dirac cone, similar to that of TR invariant TIs. The combined symmetry of translation operation and TR operation plays the role of TR symmetry in TR invariant TIs and protects the gapless nature of Dirac cones, which is similar to the previous work of R. Mong, {\it et al} \cite{mong2010}. Different from that work, our model is spinless so that the TR operation by itself satisfies $\Theta^2=1$ and cannot protect any degeneracy. The translation operation plays an essential role in inducing non-trivial topological phases, and in this sense, we dub it as ``antiferromagnetic crystalline topological insulators''. Therefore, our model indicates that spin and spin-orbit coupling are not necessary for topological phases with a single surface Dirac cone in magnetic structures.

\begin{figure}
   \begin{center}
      \includegraphics[width=3.5in,angle=0]{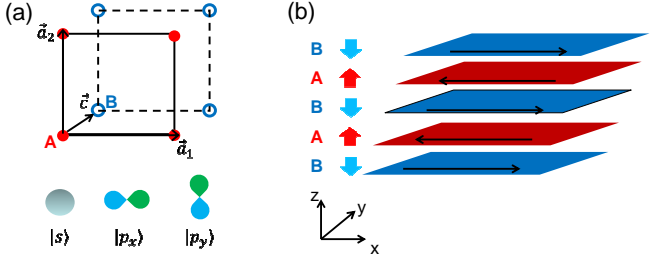}
    \end{center}
    \caption{ The schematic plot of the lattice structure. (a) For each layer, it is the square lattice with three orbitals, $s$, $p_x$ and$p_y$ orbitals. The A layer is shifted by the vector $\vec{c}$ relative to the B layer. (b) The lattice is formed by the stacking of A and B layers and each layers can be viewed as the QAH layers. 
    }
    \label{fig1}
\end{figure}

We study a model for a layered antiferromagnetic structure with each layer consisted of a square lattice with magnetic moments on each lattice site perpendicular to the xy plane, as shown in Fig. \ref{fig1}(b). We consider the case that magnetic moments in one layer are polarized along the same direction, so each layer is ferromagnetic. But the magnetic moments between two adjacent layers are polarized in the opposite directions, so the whole system has an antiferromagnetic structure with two atoms in one unit cell, denoted as A and B, as shown in Fig. \ref{fig1} (a) and (b). In one layer, there are three orbitals $|s\rangle$, $|p_x\rangle$ and $|p_y\rangle$ on each lattice site. The $|p_x\rangle$ and $|p_y\rangle$ orbitals carry the angular momentum 1 and couple to magnetic moments through Zeeman type of coupling. In one layer, the lattice vectors are denoted as $\vec{a}_1=(a,0,0)$ and $\vec{a}_2=(0,a,0)$, while layers are stacked along the direction $\vec{a}_3=2\vec{c}=(2c_1,2c_2,2c_3)$, where the vector $\vec{c}$ points from the A atom to the adjacent B atom, as shown in Fig. \ref{fig1} (a). The Hamiltonian is given by
\begin{eqnarray}
	&&H=H_{+}+H_{-}+H_{c},\label{eq:Ham1}\\
	&&H_{\eta}=\sum_{\langle \vec{n}\vec{m}\rangle_{in}, \alpha\beta}t_{\vec{n}\vec{m}}^{\alpha\beta}c^{\dag}_{\alpha \vec{n}\eta}c_{\beta \vec{m}\eta}+\sum_{\vec{n},\alpha}\epsilon_{\alpha}c^{\dag}_{\alpha \vec{n}\eta}c_{\alpha \vec{n}\eta}\nonumber\\
	&&+\sum_{\vec{n}}\eta M_1(-ic^{\dag}_{\vec{n}p_x\eta}c_{\vec{n}p_y\eta}+ic^{\dag}_{\vec{n}p_x\eta}c_{\vec{n}p_y\eta})\\
	&&H_{c}=\sum_{\langle \vec{n}\vec{m}\rangle_{c}, \alpha\beta}(r_{\vec{n}\vec{m}}^{\alpha\beta}c^{\dag}_{\alpha \vec{n}+}c_{\beta \vec{m}-}+h.c.)
\end{eqnarray}
where $\eta=+$ is for A layers and $\eta=-$ for B layers, $\vec{n}=(n_x,n_y,n_z)$, $\vec{m}=(m_x,m_y,m_z)$ denote lattice sites and $\alpha,\beta=s,p_x,p_y$ denote orbitals. $\langle nm\rangle_{in}$ represents the nearest neighbor hopping in xy plane with the hopping parameters $t^{\alpha\beta}_{\vec{n}\vec{m}}$ while $\langle \vec{n}\vec{m}\rangle_{c}$ indicates the nearest neighbor hopping between A and B atoms along $\vec{c}$ axis with the parameters $r^{\alpha\beta}_{\vec{n}\vec{m}}$. We take into account the $\sigma$ bond for s orbitals, the $\sigma$ and $\pi$ bonds for p orbitals, and the $\sigma$ bonds between s and p orbitals. For the in-plane hopping, $t^{\alpha\beta}_{nm}$ are parameterized as 
$t^{ss}_{\vec{n},\vec{n}+\hat{e}_x}=t^{ss}_{\vec{n},\vec{n}+\hat{e}_y}=t^{ss}_{\vec{n}+\hat{e}_x,\vec{n}}=t^{ss}_{\vec{n}+\hat{e}_y,\vec{n}}=u_{s\sigma}$, $t^{p_xp_x}_{\vec{n},\vec{n}+\hat{e}_x}=t^{p_yp_y}_{\vec{n},\vec{n}+\hat{e}_y}=t^{p_xp_x}_{\vec{n}+\hat{e}_x,\vec{n}}=t^{p_yp_y}_{\vec{n}+\hat{e}_y,\vec{n}}=u_{p\sigma}$, $t^{p_xp_x}_{\vec{n},\vec{n}+\hat{e}_y}=t^{p_yp_y}_{\vec{n},\vec{n}+\hat{e}_x}=t^{p_xp_x}_{\vec{n}+\hat{e}_y,\vec{n}}=t^{p_yp_y}_{\vec{n}+\hat{e}_x,\vec{n}}=u_{p\pi}$, $t^{sp_x}_{\vec{n},\vec{n}+\hat{e}_x}=t^{sp_y}_{\vec{n},\vec{n}+\hat{e}_y}=-t^{sp_x}_{\vec{n}+\hat{e}_x,\vec{n}}=-t^{sp_y}_{\vec{n}+\hat{e}_y,\vec{n}}=u_{sp\sigma}$, 
where $\hat{e}_x$ and $\hat{e}_y$ denote the vectors to the nearest neighbor site along x and y directions, respectively. For the hopping between two layers, since $\vec{c}$ is not along z direction, we need to decompose p orbitals into components along $\vec{c}$ axis and perpendicular to $\vec{c}$. Consequently, we obtain $r^{ss}_{\vec{n},\vec{n}+\hat{c}}=r^{ss}_{\vec{n}+\hat{c},\vec{n}}=v_{s\sigma}$, $r^{p_xp_x}_{\vec{n},\vec{n}+\hat{c}}=r^{p_xp_x}_{\vec{n}+\hat{c},\vec{n}}=v_{p\sigma}\lambda_1^2+v_{p\pi}(1-\lambda_1^2)$, $r^{p_yp_y}_{\vec{n},\vec{n}+\hat{c}}=r^{p_yp_y}_{\vec{n}+\hat{c},\vec{n}}=v_{p\sigma}\lambda_2^2+v_{p\pi}(1-\lambda_2^2)$,
$r^{sp_x}_{\vec{n},\vec{n}+\hat{c}}=-r^{sp_x}_{\vec{n}+\hat{e}_x,\vec{n}}=v_{sp\sigma}\lambda_1$, 
$r^{sp_y}_{\vec{n},\vec{n}+\hat{c}}=-r^{sp_y}_{\vec{n}+\hat{c},\vec{n}}=v_{sp\sigma}\lambda_2$, where $\lambda_1=\frac{c_1}{|\vec{c}|}$ ($\lambda_2=\frac{c_2}{|\vec{c}|}$) is the angle between $\vec{c}$ and x (y) axis.
$M_1$ term is the Zeeman type of coupling between p orbitals and magnetic moments. 
In the momentum space, the Hamiltonian is given by
\begin{widetext}
\begin{eqnarray}
	H_{\eta}=\sum_k\Psi^{\dag}_{\eta}\left(
	\begin{array}{ccc}
		E_s(\vec{k})&-2iu_{sp\sigma}\sin(2\pi k_1)&-2iu_{sp\sigma}\sin(2\pi k_2)\\
		2iu_{sp\sigma}\sin(2\pi k_1)&E_{px}(\vec{k})& -i\eta M_1\\
		2iu_{sp\sigma}\sin(2\pi k_2)&i\eta M_1&E_{py}(\vec{k})
	\end{array}
	\right)\Psi_{\eta}
	\label{eq:HamA}
\end{eqnarray}
\begin{eqnarray}
	H_{c}=\sum_k\Psi^{\dag}_{+}\left(
	\begin{array}{ccc}
		2v_{s\sigma}\cos(\pi k_3)&-2iv_{sp\sigma}\lambda_1\sin(\pi k_3)&-2iv_{sp\sigma}\lambda_2\sin(\pi k_3)\\
		2iv_{sp\sigma}\lambda_1\sin(\pi k_3)&2(v_{p\pi}+\lambda_1^2(v_{p\sigma}-v_{p\pi}))\cos(\pi k_3)&2(v_{p\sigma}-v_{p\pi})\lambda_1\lambda_2\cos(\pi k_3)\\
                2iv_{sp\sigma}\lambda_2\sin(\pi k_3)&2(v_{p\sigma}-v_{p\pi})\lambda_1\lambda_2\cos(\pi k_3)&2(v_{p\pi}+\lambda_2^2(v_{p\sigma}-v_{p\pi}))\cos(\pi k_3)
	\end{array}
	\right)\Psi_{-}
	\label{eq:HamAB}
\end{eqnarray}
\end{widetext}
where 
\begin{eqnarray}
	&&E_s(\vec{k})=2u_{s\sigma}(\cos(2\pi k_1)+\cos(2\pi k_2))+\epsilon_s\nonumber\\
	&&E_{px}(\vec{k})=2(u_{p\sigma}\cos(2\pi k_1)+u_{p\pi}\cos(2\pi k_2))+\epsilon_p\nonumber\\ 
	&&E_{py}(\vec{k})=2(u_{p\pi}\cos(2\pi k_1)+u_{p\sigma}\cos(2\pi k_2))+\epsilon_p. 
	\label{eq:HamEsEp}
\end{eqnarray}
Here $\Psi^\dag_{\eta}=(c^\dag_{s\eta}(\vec{k}),c^\dag_{p_x\eta}(\vec{k}),c^\dag_{p_y\eta}(\vec{k}))$, and $\vec{k}=\sum_{i=1,2,3}k_i\vec{b}_i$ with the reciprocal lattice vectors $\vec{b}_i$ satisfying $\vec{b}_i\cdot\vec{a}_j=2\pi \delta_{ij}$. The basis wavefunction for the above Hamiltonian is given by 
\begin{eqnarray}
	|\alpha\eta,\vec{k}\rangle=\frac{1}{\sqrt{N}}\sum_n e^{i\vec{k}\cdot \vec{r}_{n\eta}}\phi_{\alpha}(\vec{r}-\vec{r}_{n\eta})
	\label{eq:wf}
\end{eqnarray}
where $N$ is the normalization factor, $\vec{r}_{n\eta}=\vec{R}_n+\vec{h}_{\eta}$ with the lattice vector $\vec{R}_n$ and the position $\vec{h}_{\eta}$ for the atom A or B in one unit cell, and $\phi_\alpha$ denotes the basis wavefunction for the orbital $\alpha$ ($\alpha=s,p_x,p_y$). 

\begin{figure}
   \begin{center}
      \includegraphics[width=3.5in,angle=0]{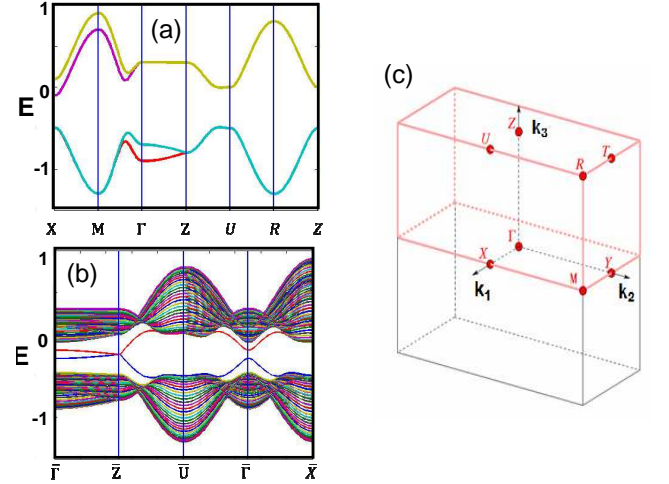}
    \end{center}
    \caption{ (a) Bulk band dispersion of the present model with the parameters given by $u_{s\sigma}=-0.2$, $u_{sp\sigma}=0.2$, $u_{p\sigma}=0.2$, $u_{p\pi}=0.2$, $v_{s\sigma}=0.05$, $v_{sp\sigma}=0.3$, $v_{p\sigma}=-0.1$, $v_{p\pi}=0.1$, $\epsilon_s=0$, $\epsilon=-5$, $M_1=4.5$, $a=1$ and $\vec{c}=(0,0,2)$. (b) The energy dispersion for a slab of the present model with the same parameters as above. A Dirac cone appears at the $\bar{Z}$ momentum. (c) The Brillouin zone of the lattice and the TR invariant momenta in the Brillouin zone.  }
    \label{fig2}
\end{figure}

When the coupling $M_1$ is large enough, each layer will be driven into the quantum anomalous Hall phase with a non-zero integer Hall conductance. To see this, we may consider one layer with the effective Hamiltonian (\ref{eq:HamA}) and change the basis from $|p_x\rangle$ and $|p_y\rangle$ to $|p_+\rangle=-\frac{1}{\sqrt{2}}(|p_x\rangle+i|p_y\rangle)$ and $|p_-\rangle=\frac{1}{\sqrt{2}}(|p_x\rangle-i|p_y\rangle)$. The effective Hamiltonian (\ref{eq:HamA}) is rewritten as 
\begin{widetext}
\begin{eqnarray}
	H_{\eta}=\left(
	\begin{array}{ccc}
		E_s(\vec{k})&\sqrt{2}iu_{sp\sigma}(\sin(2\pi k_1)+i\sin(2\pi k_2))&-\sqrt{2}iu_{sp\sigma}(\sin(2\pi k_1)-i\sin(2\pi k_2))\\
		-\sqrt{2}iu_{sp\sigma}(\sin(2\pi k_1)-i\sin(2\pi k_2))&\frac{1}{2}(E_{px}(\vec{k})+E_{py}(\vec{k}))+\eta M_1&
		-\frac{1}{2}(E_{px}(\vec{k})-E_{py}(\vec{k}))\\
		\sqrt{2}iu_{sp\sigma}(\sin(2\pi k_1)+i\sin(2\pi k_2))&-\frac{1}{2}(E_{px}(\vec{k})-E_{py}(\vec{k}))&
		\frac{1}{2}(E_{px}(\vec{k})+E_{py}(\vec{k}))-\eta M_1
	\end{array}\right). 
	\label{eq:HamA2}
\end{eqnarray}
\end{widetext}
If the parameters of this model satisfy the condition $|E_s(\vec{k})-E_p(\vec{k})-M_1|\ll 2|M_1|$ ($E_p=\frac{1}{2}(E_{px}+E_{py})$), s orbital and one of the p orbitals with the energy $E_p(\vec{k})+M_1$ are close in energy for both A and B layers, while the other p orbital are well separated. Let us assume the Fermi energy is between s orbital and the p orbital with the energy $E_p(\vec{k})+M_1$, and then the low energy physics is described by the two-band effective Hamiltonian
\begin{eqnarray}
	H_{eff,\eta}=\varepsilon(\vec{k})+\sum_{i=x,y,z}d_i\sigma_i
	\label{eq:2bHam}
\end{eqnarray}
where $\sigma$ is the Pauli matrix denoting the basis of s and p orbital. 
For A (B) layer $\eta=\pm$, $\varepsilon=\frac{1}{2}(E_s(\vec{k})+E_p(\vec{k})+M_1)$, $d_x=-\sqrt{2}u_{sp\sigma}\sin(2\pi k_2)$, 
$d_y=\mp \sqrt{2}u_{sp\sigma}\sin(2\pi k_1)$ and $d_z=\frac{1}{2}(E_s(\vec{k})-E_p(\vec{k})-M_1)$. This model has been well studied in Ref. \cite{qi2005} and it possesses a non-zero quantized Hall conductance when the configuration of the vector $\hat{d}=\frac{1}{d}(d_x,d_y,d_z)$ ($d=\sqrt{\sum_{i}d_i^2}$) in the whole Brillouin zone has a non-zero winding number. For example, we can consider the parameter regime when this model has a direct band gap around $\Gamma$ point ($\vec{k}=0$). Expanding the Hamiltonian (\ref{eq:2bHam}) around $\Gamma$ point, we find the low energy physics is described by Dirac Hamiltonian with linear dispersion. For Dirac Hamiltonian, 
a transition happens at $E_s(0)=E_p(0)+M_1$ when the band gap closes. Since the system is a normal insulator for $M_1=0$, both layers are driven into the quantum anomalous Hall phase with chiral edge states when $|M_1|>|E_s(0)-E_p(0)|$ , as depicted in Fig \ref{fig1} (b). Furthermore, because of antiferromagnetism, the Hall conductances, as well as the velocities of the corresponding chiral edge states, are opposite for two adjacent layers. Therefore, one may expect the tunneling between two layers will destroy the chiral edge states and lead to a gap opening for surface states. To test this idea, we perform numerical calculation of the energy dispersion for both the bulk and slab configuration, as shown in Fig. \ref{fig2} (a) and (b). Here we take the vector $\vec{c}=(0,0,c_3)$ and consider the side surface normal to y direction for the slab calculation. From Fig. \ref{fig2} (a), there is indeed an energy splitting for the bulk dispersion at the $\Gamma-X-Y-M$ plane (see the Brillouin zone in Fig \ref{fig2} (c)), but the energy levels at the $Z-U-R-T$ plane are still degenerate. Correspondingly, surface states open a gap at $\bar{\Gamma}$ point, but remain gapless at $\bar{Z}$ point and form a single Dirac cone in the slab configuration. Here $\bar{\Gamma}$ and $\bar{Z}$ are the projection of $\Gamma$ and $Z$ in the two-dimensional surface Brillouin zone, respectively. In the following, we will show that the gapless point of surface states at $\bar{Z}$ point is protected by the symmetry operation $T_c\Theta$, where $T_c$ is the translation operation from A layer to B layer along the vector $\vec{c}$ and $\Theta$ is TR. 

In TR invariant TIs, TR symmetry $\Theta$ plays an essential role. For spinful fermions, $\Theta=i\sigma_yK$ where $K$ is the complex conjugate and  $\sigma_y$ is the Pauli matrix for spin, so $\Theta^2=-1$, which guarantees the double degeneracy for all the states at TR invariant momenta in Brillouin zone following Kramers' theorem. This double degeneracy protects the gapless point of topological surface states. In constrast, for spinless fermions, $\Theta=K$ and $\Theta^2=1$, so TR can not protect any degeneracy by iteself. The present model has ferromagnetism within one layer and antiferromagnetism between two adjacent layers, so the Hamiltonian has a combined symmetry $T_c\Theta$. To understand the role of $T_c\Theta$, we may calculate the behavior of $(T_c\Theta)^2$ on the basis (\ref{eq:wf}) and find
\begin{eqnarray}
	&&(T_c\Theta)^2|\alpha\eta,\vec{k}\rangle=(T_c\Theta)e^{-i\vec{k}\cdot\vec{c}}|\alpha\bar{\eta},-\vec{k}\rangle\nonumber\\
	&&=e^{i\vec{k}\cdot\vec{a}_3}|\alpha\eta,\vec{k}\rangle
	\label{eq:TcTheta2}
\end{eqnarray}
where $\bar{\eta}=-\eta$ means that A and B layers are interchanged. Thus, $(T_c\Theta)^2=e^{i2\pi k_3}$ depends on the value of $k_3$. For the plane $k_3=0$, $(T_c\Theta)^2=1$, but for the plane $k_3=\frac{1}{2}$, $(T_c\Theta)^2=-1$, which explains the doubly degenerate states at TR invariant momenta when $k_3=\frac{1}{2}$. In Fig \ref{fig2} (a), we find there is a double degeneracy for each momentum in the $k_3=\frac{1}{2}$ plane and this is because we take $\vec{c}=(0,0,c_3)$ along z axis and the Hamiltonian has additional two-fold rotation symmetry $C_2$ around z axis. On the basis $|\alpha\eta,\vec{k}\rangle$, the $C_2$ operation is written as $C_2|\alpha\eta,\vec{\kappa},k_3\rangle=\nu_\alpha|\alpha\eta,-\vec{\kappa},k_3\rangle$, where $\vec{\kappa}=(k_1,k_2)$ and $\nu_{\alpha}$ is 1 for s orbitals and -1 for p orbitals. Therefore, for any eigen state $|\Psi_m(\vec{\kappa})\rangle$ in the $\kappa=\frac{1}{2}$ plane, where $m$ denotes different eigen states, $C_2T_c\Theta|\Psi_m(\vec{\kappa})\rangle$ is also an eigen state with the same eigen energy at the same momentum. Moreover, when $k_3=\frac{1}{2}$, $T_c\Theta|\alpha\pm,\vec{\kappa}\rangle=e^{-\frac{i}{2}\vec{b}_3\cdot\vec{c}+i\vec{b}_3\cdot\vec{h}_{\bar{\eta}}}|\alpha\mp,-\vec{\kappa}\rangle=\mp i|\alpha\mp,-\vec{\kappa}\rangle$, so one can see that the operation $T_c\Theta$ at the $k_3=\frac{1}{2}$ plane is quite similar to TR operation for spinful fermions. Consequently, one can show that $\langle\Psi_m(\vec{\kappa})|C_2T_c\Theta|\Psi_m(\vec{\kappa})\rangle=\langle\Theta C_2T_c\Theta\Psi_m(\vec{\kappa})|\Theta\Psi_m(\vec{\kappa})\rangle=-\langle\Psi_m(\vec{\kappa})|C_2T_c\Theta|\Psi_m(\vec{\kappa})\rangle=0$, so the eigen states $C_2T_c\Theta|\Psi_m(\vec{\kappa})\rangle$ and $|\Psi_m(\vec{\kappa})\rangle$ are orthogonal and there is a double degeneracy for each momentum $\vec{\kappa}$ at $k_3=\frac{1}{2}$ plane.   

\begin{figure}
   \begin{center}
      \includegraphics[width=3.5in,angle=0]{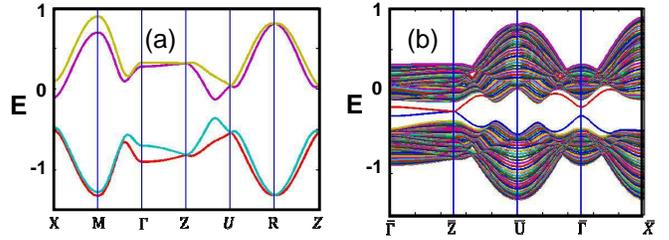}
    \end{center}
    \caption{ (a) Bulk band dispersion with the lattice vectors $a=1$ and $\vec{c}=(0.1,0.15,2)$. All the other parameters are the same as above. The degeneracies at Z, U, R, T are preserved due to the $T_c\Theta$ symmetry. (b) Energy dispersion for a slab of the present model.     }
    \label{fig3}
\end{figure}

When $c_{1,2}$ in $\vec{c}$ is finite, $C_2$ rotation symmetry is broken, so the degeneracy at $k_3=\frac{1}{2}$ plane is removed, except four momenta $Z$, $U$, $R$ and $T$, as shown in Fig \ref{fig3} (a). These momenta are invariant under $T_c\Theta$, so $T_c\Theta$ by itself protects the degeneracy. One can directly show $\langle\Psi_m(\Lambda)|T_c\Theta|\Psi_m(\Lambda)\rangle=-\langle\Psi_m(\Lambda)|T_c\Theta|\Psi_m(\Lambda)\rangle=0$ with $\Lambda=Z, U, R, T$, so $|\Psi_m(\Lambda)\rangle$ and $T_c\Theta|\Psi_m(\Lambda)\rangle$ give two degenerate eigen states. 
From Fig. \ref{fig3} (b), one finds that the gapless Dirac point of surface states at $\bar{Z}$ also remains. 


The $Z_2$ topological invariant can be defined in the $k_3=\frac{1}{2}$ plane due to the degeneracy at $Z, U, R, T$. Since we have only four $T_c\Theta$-invariant momenta, only one $Z_2$ number is allowed, and given by\cite{fu2007a}
\begin{eqnarray}
	(-1)^\nu=\Pi_i\delta_i,\qquad \delta_i=\frac{\sqrt{det[w(\Lambda_i)]}}{Pf[w(\Lambda_i)]} 
	\label{eq:Z2}
\end{eqnarray}
where $\nu$ is $Z_2$ topological number and $\Lambda_i$ is taken to be $Z$, $U$, $R$, $T$. 
The matrix $w$ is defined as $w_{mn}(\vec{\kappa})=\langle \Psi_m(-\vec{\kappa})|T_c\Theta|\Psi_n(\vec{\kappa})\rangle$ with $m$ and $n$ taken from all the occupied bands. $det[w(\Lambda_i)]$ and $Pf[w(\Lambda_i)]$ are the determinant and pfaffian of $w$ at $\Lambda_i$. Alternatively, one can also define $Z_2$ invariant with the Berry gauge potential $A_{x,y}=\sum_{m}\langle \Psi_m|\frac{\partial}{\partial k_{x,y}}|\Psi_m\rangle$ and the Berry curvature $F_{xy}=\frac{\partial}{\partial k_x}A_y-\frac{\partial}{\partial k_y}A_x$, where $m$ is for all the occupied bands. The $Z_2$ number is given by\cite{fu2006}
\begin{eqnarray}
	\nu=\frac{1}{2\pi}\left[ \oint_{\partial \Omega_{1/2}}d\vec{\kappa}\cdot\vec{A}-\int_{\Omega_{1/2}} d^2\kappa F_{xy}\right] \mbox{mod} 2
	\label{eq:Z22}
\end{eqnarray}
where $\Omega_{1/2}$ takes half of the $k_3=\frac{1}{2}$ plane. The expression (\ref{eq:Z22}) is much easier to evaluate numerically and with the numerical method introduced in Ref. \cite{fukui2007a}, we obtain $\nu=1$ for the parameters given in the caption of Fig. \ref{fig2} and \ref{fig3}.  

For TR invariant TIs, when there is an inversion symmetry, $Z_2$ invariant can be easily determined by the parity eigenvalue of the occupied bands\cite{fu2007a}. However, this simplification cannot be applied here because inversion $P$ does not commutate with $T_c\Theta$, $PT_c\Theta=T_{-c}\Theta P\neq T_c\Theta P$. It turns out that $C_2$ symmetry, which commutates with $T_c\Theta$, plays the role of inversion symmetry in TR invariant TIs. If there is $C_2$ symmetry in our model, which requires the vector $\vec{c}$ along z direction, $Z_2$ invariant can be calculated by the eigenvalues of $C_2$ rotation of all the occupied bands, with the expression\cite{fu2007a}
\begin{eqnarray}
	(-1)^\nu=\Pi_i\delta_i,\qquad \delta_i=\Pi_{m=1}^N \xi_{2m}(\Lambda_i)
	\label{eq:Z23}
\end{eqnarray}
where $\xi_{2m}(\Lambda_i)=\pm1$ is the $C_2$ eigenvalue of the $2m$th occupied band. The total number of the occupied bands is $2N$. Since the $(2m-1)$th and $2m$th bands are related to each other by $T_c\Theta$, these two states share the same $C_2$ eigenvalue, $\xi_{2m}(\Lambda_i)=\xi_{2m-1}(\Lambda_i)$. For our present model, s orbitals are even under $C_2$ rotation while p orbitals are odd. Therefore, topological phase transition happens when there is a band inversion between s orbital and one of the p orbitals at high symmetry momenta. This is consistent with the above analysis based on the two-band effective model. With the parameters given in the caption of Fig \ref{fig2}, one find $\nu=1$ for our model, which confirms the topological nature of gapless surface states.


The model presented here is quite similar to the antiferromagnetic model in Ref. \cite{mong2010}, but there is one essential difference. Ref. \cite{mong2010} discusses the model of spinful fermions with the TR operator $\Theta=i\sigma_yK$, so $\Theta^2=-1$ by itself. Our model is for spinless fermions, so $\Theta^2=1$ and the degeneracy originates from the combination of time reversal and translation operation. Consequently, in Ref. \cite{mong2010}, $(T_c\Theta)^2=-1$ for $k_3=0$ plane and $(T_c\Theta)^2=1$ for $k_3=\frac{1}{2}$ plane, yielding the $Z_2$ topological invariant only well-defined at $k_3=0$ plane. It is exactly opposite in our case. In Ref.\cite{mong2010}, for the $Z_2$ topological invariant at the $k_3=0$ plane, there is no difference between inversion $P$ and two-fold rotation $C_2$ along z axis. But in our case, these two operations are different for the $k_3=\frac{1}{2}$ plane, and only $C_2$ rotation can be used to define $Z_2$ invariant. Since $Z_2$ invariants are defined at different momentum planes for two cases, the corresponding surface Dirac cones also appear at different high symmetry momenta. Our model shows that for magnetic systems, topological phases with a single surface Dirac cone can exist in a spinless fermion model.

In conclusion, we study a spinless model in an antiferromagnetic structure, which has the topologically non-trivial phase with a single surface Dirac cone. The bulk $Z_2$ invariants are evaluated and compared with the direct numerical calculation of surface states in a slab configuration. Our model indicates the possibility to generalize the concept of TIs to magnetic structures\cite{dresselhaus2008} and other boson systems. 

We would like to thank C. Fang, X. Liu, Y.H. Wu for useful discussions. 


%

%

\end{document}